\documentclass[preprint2]{aa}
\usepackage[utf8]{inputenc}
\usepackage[varg]{txfonts}
\usepackage{graphicx}
\usepackage{natbib}
\usepackage{stfloats}
\bibpunct{(}{)}{;}{a}{}{,}
\usepackage{multirow}
\usepackage{amsmath}
\usepackage[normalem]{ulem}
\usepackage[colorlinks=true, allcolors=blue]{hyperref}

\newcommand{\be}{\begin{equation}}
\newcommand{\ee}{\end{equation}}
\newcommand{\ud}{\rm d}

\begin{document}



\title{SOLIS XVII: Jet candidate unveiled in OMC-2 and\\ its possible link to the enhanced cosmic-ray ionisation rate\thanks{Based on observations carried out with the IRAM NOEMA interferometer. IRAM is supported by INSU/CNRS (France), MPG (Germany), and IGN (Spain).}}


\author{V. Lattanzi\inst{1}, F.\,O. Alves\inst{1}, M . Padovani\inst{2}, F. Fontani\inst{2}, P. Caselli\inst{1}, C. Ceccarelli \inst{3}, A. L\'{o}pez-Sepulcre\inst{3,4}, C. Favre\inst{3}, R. Neri\inst{4}, L. Chahine\inst{4,5}, C.Vastel\inst{6}, \and L. Evans\inst{2,6}}

\institute{Center for Astrochemical Studies, Max-Planck-Institut f\"{u}r Extraterrestrische Physik, Gießenbachstraße~1, 85748 Garching, Germany \and INAF-Osservatorio Astrofisico di Arcetri, Largo E. Fermi 5, 50125 Firenze, Italy \and Univ. Grenoble Alpes, CNRS, IPAG, F-38000 Grenoble, France \and Institut de Radioastronomie Millimetrique, 300 rue de la Piscine, Domaine Universitaire de Grenoble, 38406, Saint-Martin d’H\`{e}res, France \and \'Ecole doctorale de Physique, Universit\'e Grenoble Alpes, 110 Rue de la Chimie, 38400 Saint-Martin-d'H\`eres, France \and IRAP, Universit\'{e} de Toulouse, 9 avenue du colonel Roche, 31028 Toulouse Cedex 4, France\\ \email{lattanzi@mpe.mpg.de}}

\abstract{
The study of the early phases of star and planet formation is important to understand the physical and chemical history of stellar systems such as our own. In particular, protostars born in rich clusters are prototypes of the young Solar System.}
{In the framework of the Seeds Of Life In Space (SOLIS) large observational project, the aim of the present work is to investigate the origin of the previously inferred high flux of energetic particles in the protocluster FIR4 of the Orion Molecular Cloud 2 (OMC-2), which appears asymmetric within the protocluster itself.
}
{Interferometric observations carried out with the IRAM NOEMA interferometer were used to map the silicon monoxide (SiO) emission around the FIR4 protocluster. Complementary archival data from the ALMA interferometer were also employed to help constrain excitation conditions. A physical-chemical model was implemented to characterise the particle acceleration along the protostellar jet candidate, along with a non-LTE analysis of the SiO emission along the jet.}
{The emission morphology of the SiO rotational transitions hints for the first time at the presence of a collimated jet originating very close to the brightest protostar in the cluster, HOPS-108. 
}
{The NOEMA observations unveiled a possible jet in the OMC-2 FIR4 protocluster propagating towards a previously
measured enhanced cosmic-ray ionisation rate. This suggests that energetic particle acceleration by the jet shock close to the protostar might be at the origin of the enhanced cosmic-ray ionisation rate, as confirmed by modelling the protostellar jet.   
}


\authorrunning{Lattanzi et al.}\titlerunning{OMC-2 FIR4 with NOEMA}

\maketitle

\section{Introduction}

The early phases of star formation are known to be highly dynamic, with accretion of material from the surrounding envelope to the protostar and, at the same time, energetic outflows that contribute to the dispersal of the mother cloud. This phase is also associated with a very rich chemical and physical complexity, which has been reported extensively in observational works \citep[e.g.][]{1538-4357-593-1-L51, 2005ApJ...632..973J}. Many studies conducted in the past few decades propose that stars generally do not form in isolation. \citet{2010ARA&A..48...47A} showed that our Solar System is likely to have formed in a moderately large cluster environment. Moreover, analysis of the short-lived radionuclides within meteoritic material indicates that during its early evolutionary phases, our Sun experienced a high flux of energetic ($\geq$10\,MeV) particles \citep{2013ApJ...763L..33G}.

The region of the Orion Molecular Cloud north of the Orion Nebula, and known as OMC-2 FIR4, is the closest prototype of an intermediate-to-high-mass protocluster. The short distance to the Solar System \citep[388$\pm$5\,pc,][]{2017ApJ...834..142K} allows for a detailed view of its structure through high spatial resolution observations. Several studies focused on the structure and chemistry of this source identified at least six compact continuum sources in the millimetre and submillimetre.
Observations carried out with the \textit{Herschel} space telescope recorded the first indirect evidence of an enhancement of energetic particles in the protocluster FIR4 \citep{Ceccarelli+2014}. More recently, \citet{Fontani+2017} mapped the different distribution of HC$_5$N and HC$_3$N towards FIR4. The spatial differentiation between the two cyanopolyynes indicates a higher cosmic-ray (CR) ionisation rate (3 orders of magnitude larger than the average interstellar value, $\zeta \approx 10^{-17}$ s$^{-1}$) in the eastern part of the region, where HC$_5$N peaks.  These findings were confirmed by \citet{Favre+2018}, who mapped the excitation temperature across the FIR4 region using observations of c-C$_3$H$_2$. With the aid of chemical modelling including photodissociation, the observational data can be reproduced only assuming a high CR ionisation rate ($\zeta \approx$ 4$\times$10$^{-14}$ s$^{-1}$), very similar to the one constrained by the \citet{Ceccarelli+2014} analysis. OMC-2 FIR4 is thus considered one of the best analogues of our Solar System progenitor \citep{Favre+2018, Fontani+2017}. Nevertheless, although it is now clear from the previous analyses that FIR4 is permeated by a flux of highly energetic CR-like ionising particles, less is known about their origin, which has to be internal \citep{Ceccarelli+2014}. 

The findings of an enhanced ionisation rate has also generated great interest from a theoretical point of view, as models have been showing that thermally charged particles can be accelerated in shock fronts along jets driven by young stellar objects, according to the first-order Fermi acceleration mechanism \citep{Padovani+2015,Padovani+2016,GachesOffner2018,Padovani+2021}. These energetic particles can thus explain the high CR ionisation rate estimated by observations. However, the origin of the increased ionisation rate towards the east side of the region has not yet being identified.

Due to the interest of the source and its vicinity, many other studies investigated the complexity of the OMC-2 region. \citet{Gonzalez-Garcia+_2016} showed, using \textit{Herschel/}PACS observations, a [O I] jet originating from FIR3 and connecting FIR3 to FIR4. Multi-wavelength and multi-epoch VLA observations from \citet{Osorio+2017} resolved a collimated synchrotron emission following a similar morphology to the jet observed by \citet{Gonzalez-Garcia+_2016}. The interaction of this non-thermal jet emitting from HOPS-370 (i.e. FIR3) and the surrounding material in FIR4 is proposed by the authors as the formation mechanism of HOPS-108, as previously suggested by \citet{Shimajiri+_2008}.

Being a prototype of the young Solar nebula, OMC-2 FIR4 is one of the targets in the Seeds Of Life In Space \citep[SOLIS;][]{2017ApJ...850..176C} large programme. Through interferometric observations carried out with the IRAM NOrthern Extended Millimeter Array (NOEMA) at different frequencies and antenna configurations, the goal of the SOLIS project is to understand how molecular complexity grows in Solar-type star forming regions. The aim of the present work is to show initial evidence of a jet source within the protocluster and associated with the brightest protostar in FIR4, HOPS-108, a hot corino with a luminosity of $\sim37~L_\odot$ \citep{Furlan+_2014,Tobin+2019,Chahine+_2022}. The emission of the $J$\,=\,2-1 rotational transition of silicon monoxide, SiO, a well-known shock tracer within star forming regions, is used as a kinematic tracer. 
The same transition was observed by \citet{Shimajiri+_2008} with the Nobeyama Millimeter Array, although only the emission from the brightest compact region in the eastern part of FIR4 was mapped due to the sensitivity of the observations.
To complement our analysis, we used SiO $J$\,=\,5-4 archival data from the ALMA telescope (project 2017.1.01353.S, PI: S. Takahashi), which include observations with both the main (12-m antennas) and the compact (7-m antennas) array. A full detailed analysis of the ALMA observations was described recently in \citet{2022arXiv221112140S}. 

The observational setup is presented in the following section. In Sect.\,3, a description of the analysis and the obtained results is provided. The model for the particle acceleration is also presented therein before the discussion (Sect.\,4). The main outcomes of the work are described in the last section.

\section{Observations}\label{sec:obs}

\subsection{NOEMA}

The interferometric observations were carried out in several runs in 2016 and 2017. The IRAM NOEMA array was used in C configuration as part of the SOLIS large programme \citep{2017ApJ...850..176C}. All data reported here were obtained with the WideX band correlator, which provides 1843 channels over 3.6\,GHz of bandwidth with a channel width of 1.95\,MHz ($\sim$\,6.5\,km\,s$^{-1}$ at 86\,GHz). The phase centre of the observations was RA(J2000) = $05^{h}35^{m}26\fs97$, DEC(J2000) = $-05\degr09\arcmin56\farcs8$, and the local standard of rest velocity was set to 11.4\,km\,s$^{-1}$, and the systemic velocity was OMC-2 FIR4 \citep{2015ApJS..221...31S,Favre+2018}. The primary beam is $\sim$\,54$^{\prime\prime}$ at 86\,GHz, and the system temperature during the observations ranged from 60--100\,K ($\sim$\,200\,K in summer) with an amount of precipitable water of $\lesssim$\,5\,mm (10--15\,mm in summer). The maximum recoverable scale is 20\,arcsec. The absolute flux scale was calibrated by observing the quasars LKHA101 and MWC349, while 3C454.3 and 3C84 were used as a calibrator for the bandpass shape. For gain (phase and amplitude) calibration, 0414-189, 2200+420, 0524+034, and 0539-057 were used.

Calibration and imaging were performed using the CLIC and MAPPING software of the GILDAS package\footnote{\url{http://www.iram.fr/IRAMFR/GILDAS/}}, respectively. The continuum was imaged by averaging the line-free channels of the WideX backend. The continuum image was self-calibrated and the solutions were applied to the spectral lines. A natural weight was used in the visibilities, and all the cleaning of the detected spectral features was performed using the H\"{o}gbom method \citep{1974A&AS...15..417H}. All the maps presented in this work are primary beam corrected, and the final synthesised beam is 3$\farcs$1$\times$1$\farcs$4 (P.A.= $-$161$\degr$) at 86\,GHz (see Tables\,\ref{table:1} and \ref{table:2}). \\

\subsection{ALMA}

Additional SiO emission data from  OMC-2 FIR4 were obtained from Atacama Large Millimeter/`submillimeter Array (ALMA) observations in the frame of the 2017.1.01353.S project (PI: S. Takahashi), with an observing run in April 2018 for a total time on source of $\sim$\,30\,minutes. A total of 44 antennas were used for the observations in Band 6 (1.3\,mm); flux and bandpass calibration were obtained through observations of J0522-3627, while the quasar J0541-0541 was used for phase and amplitude gain calibration. The shortest and longest projected baselines are 15\,m and 500\,m, respectively, with a maximum recoverable scale of $\sim$\,11\,arcsec. The data were processed and primary beam corrected using standard ALMA calibration scripts of the Common Astronomy Software Applications (CASA\footnote{CASA is developed by an international consortium of scientists based at the National Radio Astronomical Observatory (NRAO), the European Southern Observatory (ESO), the National Astronomical Observatory of Japan (NAOJ), the Academia Sinica Institute of Astronomy and Astrophysics (ASIAA), the CSIRO division for Astronomy and Space Science (CASS), and the Netherlands Institute for Radio Astronomy (ASTRON) under the guidance of NRAO.}, version 5.4.0) package. The final synthesised beam of the SiO $J$\,=\,5-4 map is 1$\farcs$2$\times$0$\farcs$7 (P.A.= -68$\degr$) at 217\,GHz (see Table\,\ref{table:1}).

\begin{table}\small
\caption{Observational parameters.}
\label{table:1} \centering
\begin{tabular}{ccccc}
\hline\hline   
Setup &  Freq. range & Chan. width & Synth. Beam     & P.A.  \\
      &    [GHz]         &    [MHz]      &  [$^{\prime\prime}$] & [${\degr}$]   \\
\hline
NOEMA  & 83.680\,--\,87.280    &  1.95   & 3.1\,$\times$\,1.4  & $-161$  \\
ALMA    & 216.598\,--\,217.598  & 0.24    & 1.2\,$\times$\,0.7  &  $-68$  \\
\hline
\end{tabular}
\end{table}

\begin{table}
\caption{Main spectroscopic properties of the SiO lines.}
\label{table:2} \centering
\begin{tabular}{cr@{.}lcc}
\hline \hline
Transition & \multicolumn{2}{c}{Frequency} & E$_{\rm up}$/k$_{\rm b}$ & A$_{ul}$  \\
           & \multicolumn{2}{c}{[MHz]}     & [K]  & [s$^{-1}$]\\
\hline 
$J$\,=\,2--1  &   86846 & 985  &  6.3 &   2.93~$\,\times\,$~10$^{-5}$ \\
$J$\,=\,5--4  &  217104 & 919  & 31.3 &   5.20~$\,\times\,$~10$^{-4}$ \\

\hline 
\end{tabular}
\tablefoot{Spectroscopic data and parameters from \citet{2013JPCA..11713843M} and available in the Cologne Database for Molecular Spectroscopy molecular line catalogue \citep[CDMS;][]{2016JMoSp.327...95E}. E$_{\rm up}$/k$_{\rm b}$ is the energy of the upper state in Kelvin, k$_{\rm b}$ the Boltzmann constant, and A$_{ul}$ is the Einstein coefficient for spontaneous emission in s$^{-1}$.} 
\end{table}

\section{Results}

\subsection{Morphology of the SiO emission}\label{sec:morphology}

\begin{figure}[htbp]
    \centering
    \includegraphics[width=\columnwidth]{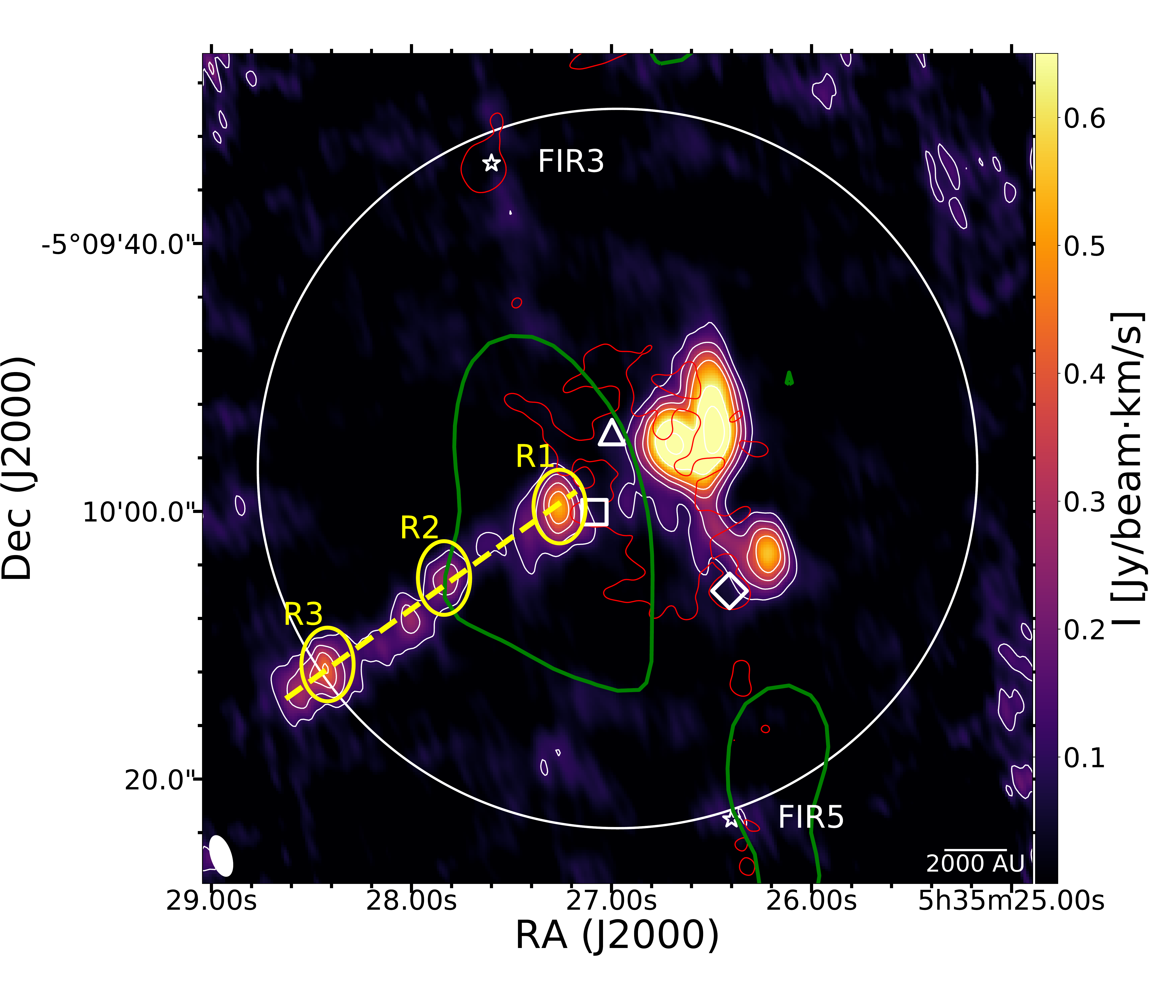}
    \caption{Integrated intensity (between $-18$ and $43$\,km\,s$^{-1}$) map of the SiO ($J$\,=\,2--1) emission towards OMC-2 FIR4; white contours are 10\%, 20\%, 30\%, 40\%, 60\%, and 80\% of the maximum value (0.985\,Jy/beam$\cdot$km\,s$^{-1}$). The velocity channels were selected in the emission map with a 5$\sigma$ cut-off around the emission peak (1$\sigma=6.7\times10^{-4}$\,Jy/beam). Red contour indicates the 10$\sigma$ of  the 85\,GHz continuum emission (Neri et al. in preparation; $\sigma= 3.8\,\times\,10^{-5}$ Jy/beam). Green contour is the 7.5$\sigma$ integrated emission of HC$_5$N described in \citet{Fontani+2017}, with 1$\sigma = 3.6\times10^{-3}$Jy/beam$\cdot$km\,s$^{-1}$. This region, with an average radius of $\sim$\,5000\,AU, is also where \citet{Fontani+2017} inferred the CR ionisation rate of $\zeta$=4$\times$10$^{-14}$\,s$^{-1}$. The yellow open ellipses along the jet represent the three regions used to extract the spectra shown in Fig.\,\ref{fig:SiO_spectra} and are labelled R1, R2, and R3, while the yellow dashed line encompassing these three regions shows the cut used to generate the PV plot in Fig.\,\ref{fig:pv_plot}.  The three main protostars in FIR4, namely HOPS-108 (RA(J2000) = $05^{h}35^{m}27\fs086$, DEC(J2000) = $-05\degr10\arcmin00\farcs06$), HOPS-64 (RA(J2000) = $05^{h}35^{m}26\fs998$, DEC(J2000) = $-05\degr09\arcmin54\farcs08$), and VLA15 (RA(J2000) = $05^{h}35^{m}26\fs41$, DEC(J2000) = $-05\degr10\arcmin05\farcs94$) \citep{Tobin+2019}, are represented by the three open symbols: square, triangle, and diamond, respectively. The white open circle marks the primary beam Full Width Half Maximum (FWHM) for the NOEMA observations. The bottom left ellipse represents the synthesised beam.}
    \label{fig:SiO_mom0}
\end{figure}

Figure\,\ref{fig:SiO_mom0} shows the integrated intensity of the SiO $J$\,=\,2--1 emission. A 5$\sigma$ intensity cut, with 1$\sigma$ at 6.7$\times$10$^{-4}$\,Jy/beam measured before primary beam correction, was adopted to select the channels with significant emission for the integration  ($-18<v_{lsr}<43$~km\,s$^{-1}$).

\begin{figure}[htbp]
    \includegraphics[width=\columnwidth]{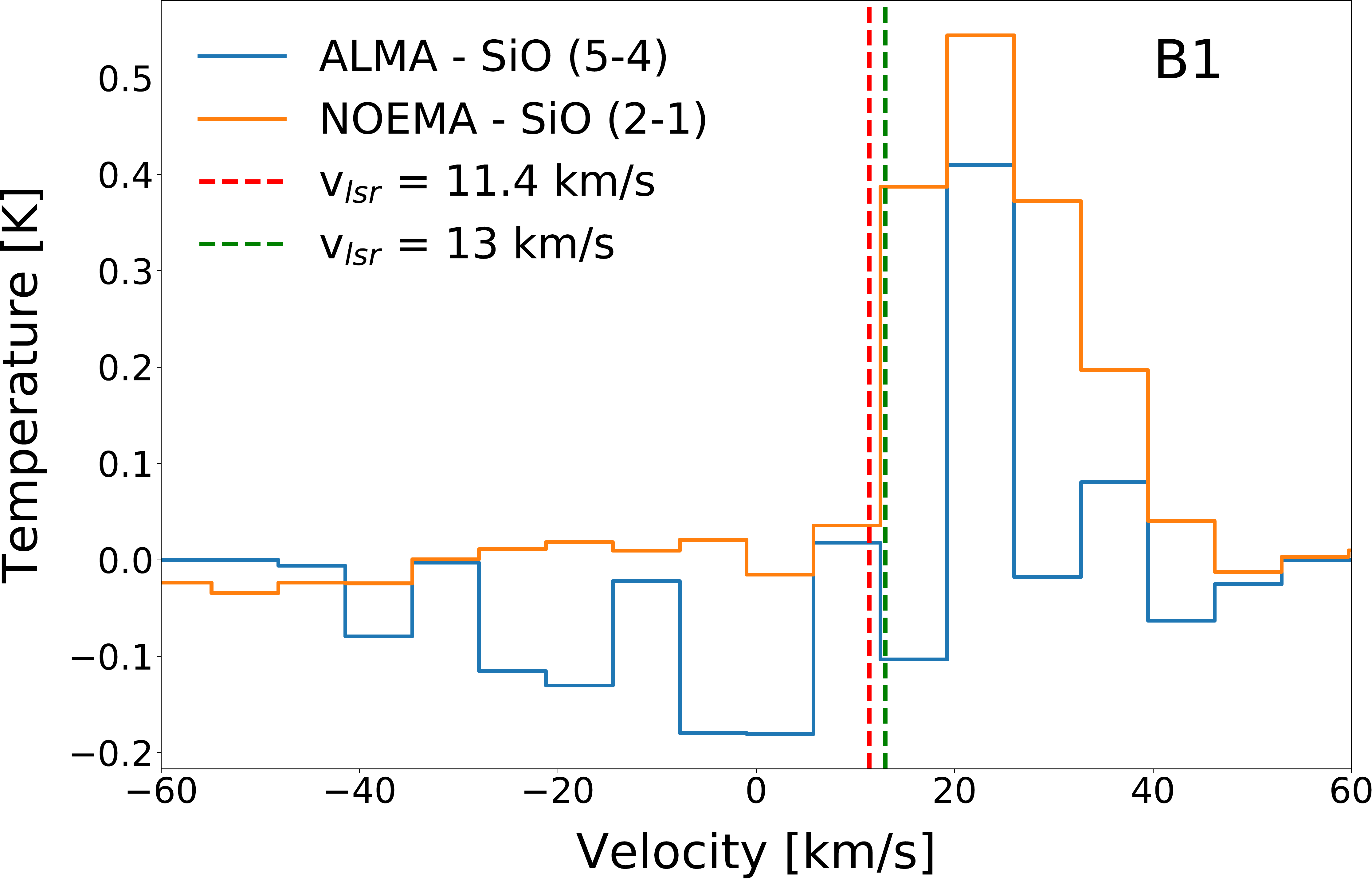}\\
    \includegraphics[width=\columnwidth]{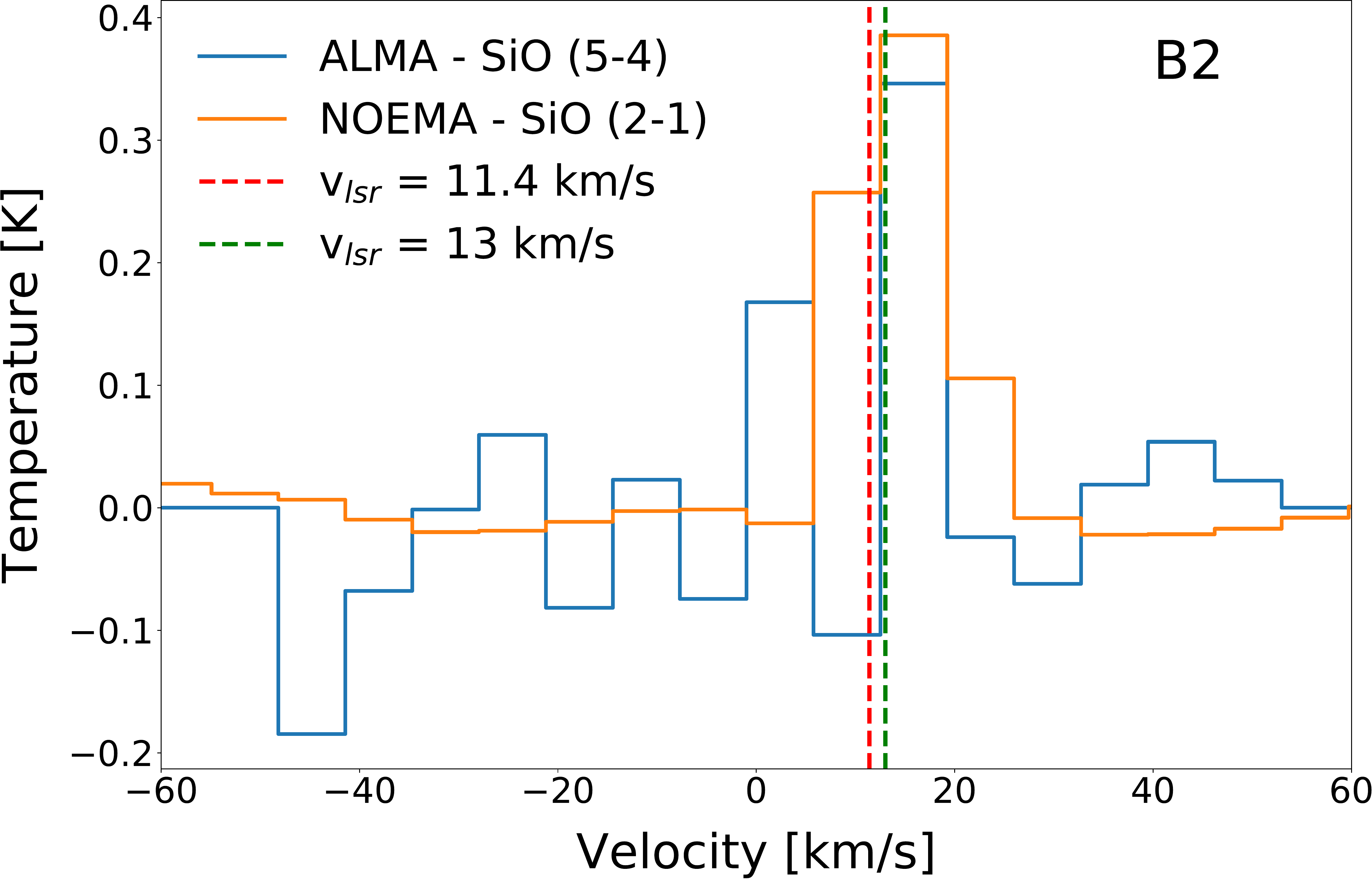}\\
    \includegraphics[width=\columnwidth]{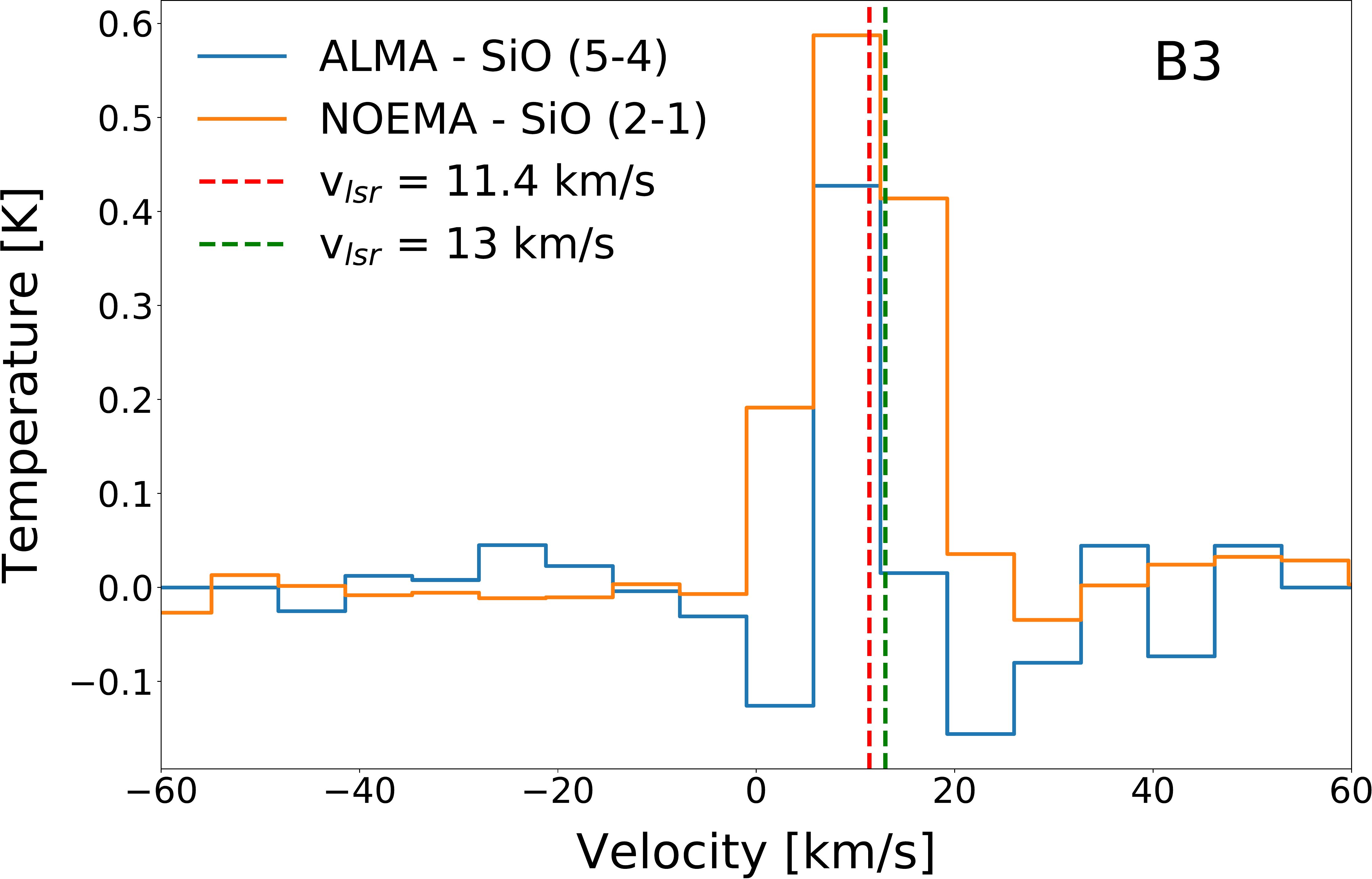}
    \caption{SiO emission line spectra extracted from the three regions described in Fig.\,\ref{fig:SiO_mom0}. The red and green dashed vertical lines correspond to the systemic velocity of the FIR4 envelope and the protostar HOPS-108 at 11.4\,km\,s$^{-1}$ and 13\,km\,s$^{-1}$, respectively. The ALMA spectra were obtained after performing a convolution of the ALMA beam with a Gaussian beam as large as the NOEMA angular resolution and regridding it to the same pixel size.}
    \label{fig:SiO_spectra}
\end{figure}

\noindent The morphology of the SiO 2-1 emission can be described as two main blocks with quite different morphologies. The west component is compact and extends in a north-south direction, with the brightest region in the northern part, a few thousand AU west of the HOPS-64 protostar, which was previously detected in near- and mid-infrared bands by \textit{Herschel} \citep{2012ApJ...749L..24A, Furlan+_2014}. A second peak of the emission can be seen at a similar distance north-west of the radio source VLA15 \citep{Osorio+2017}. In contrast, the emission on the eastern side of FIR4 is clumpy and filamentary, tending towards a north-west south-east direction. The brightest clump is located very close ($\la$\,1000\,AU) to the HOPS-108 protostar. \\
\indent The connection of this filamentary emission and the nearby protostar HOPS-108 is the focus of the present work and is described in more detail in the following. The brightest emission on the western part of the source, which was also observed in SiO emission by \citet{Shimajiri+_2008}, is not the subject of our present analysis. As recently shown by \citet{2022A&A...667A...6C}, this part of FIR4 requires angular and spectral resolution analyses to disentangle the several jets propagating from the members of the protocluster system.\\ 
\indent Sample spectra were extracted along the collimated emission for a deeper analysis. The spectra are shown in Fig.\,\ref{fig:SiO_spectra} and were extracted from the NOEMA and ALMA maps from three regions\footnote{The negative lobes in the ALMA spectra, especially in R1 and R3, are likely due to the filtering of larger scale SiO emission of the ALMA interferometer. These effects are present both the in the 12m array and in the compact array data, and both configurations have a maximum recovery scale smaller than the NOEMA array, whose data do not show this effect. If somehow this effect impacts the line width of the derived spectra, and in part its absolute brightness peak, this information is not the focus of the present work; only the latter quantities were used, with substantial uncertainties, in the non-LTE analysis. The purpose of the displayed spectra here is to show the kinematics of the SiO peaks at the different locations, i.e. the three regions along the jet emission.}: the bright spot (6$\sigma$ level) map close to the HOPS-108 protostar and two regions encompassing emission knots chosen along the jet (yellow ellipses in Fig.\,\ref{fig:SiO_mom0}). The region close to the protostar was labelled R1 (in Fig.\,\ref{fig:SiO_mom0} and Fig.\,\ref{fig:SiO_spectra}) and the farthest R3, with the R2 being halfway between those two. For consistency, the ALMA maps around the SiO 5-4 emission line, exhibiting the same morphology of the SiO 2-1 emission, were convolved with a 2D Gaussian beam to match NOEMA angular resolution and regridded to the same pixel size. The velocity resolution of the ALMA data was also binned to match the NOEMA one ($\sim 7$ km\,s$^{-1}$). Despite the low velocity resolution, it is clear from Fig.\,\ref{fig:SiO_spectra} that the R1 emission is at a higher velocity with respect to the systemic velocity of the envelope (11.4\,km\,s$^{-1}$) and to the HOPS-108 protostar  \citep[13\,km\,s$^{-1}$,][]{Tobin+2019}.

Further considerations on the kinematics can be drawn by studying the emission distribution for different channels. Fig.\,\ref{fig:SiO_channels} shows that the SiO emission is detected through velocity channels from $-14.5$ to $32.7$\,km\,s$^{-1}$. The emission at velocities higher than the HOPS-108 systemic velocity, $\sim$\,13\,km\,s$^{-1}$ \citep{Tobin+2019}, remains very close to the protostar itself and coincides with the R1 location. Going from HOPS-108 to R3, the SiO emission defines a mild S-shape jet moving from the protostar towards the edge of the FIR4 cloud. 
In this picture, the SiO `blue' outflow looks projected in the plane of the sky, while the red lobe is probably mixed with the complexity of the western region of FIR4 where HOPS-64 and VLA15 are also present. In this case, the high-velocity blob near the HOPS-108 protostar might be a product of the synchrotron emission of FIR3 impacting the region in FIR4, or, more likely, this could be due to the mixing of the various jets driven by the YSOs within the protocluster. Recent 100-au scale analysis by \citet{2022A&A...667A...6C} showed a complex, filamentary structure of the western region of FIR4 at high resolution. In particular, SiO emission unveiled the presence of multiple bow-shock features with sizes between $\sim$\,500 and 2700 au likely caused by a precessing jet from FIR3 that goes from east to west. 

\begin{figure}[htbp]
    \centering
    \includegraphics[width=\columnwidth]{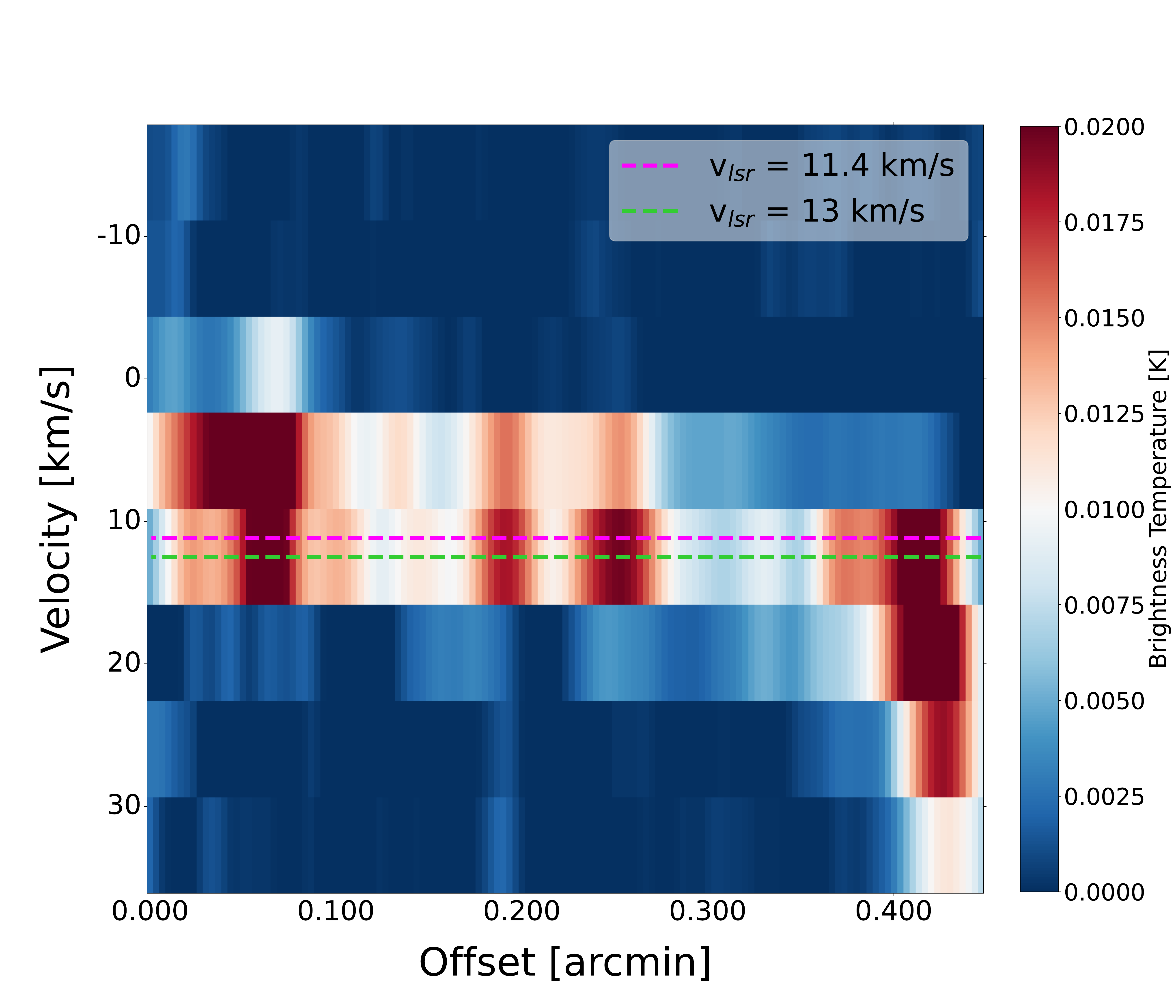}
    \caption{Position-velocity plot obtained along the cut showed in Fig.\,\ref{fig:SiO_mom0}. The zero offset position is the south-easternmost point of the cut in Fig.\,\ref{fig:SiO_mom0}. Magenta and green dashed lines represent the velocity of the embedding cloud and HOPS-108 protostar, respectively.}
    \label{fig:pv_plot}
\end{figure}

\begin{figure*}[h]
    \centering
    \includegraphics[width=\textwidth]{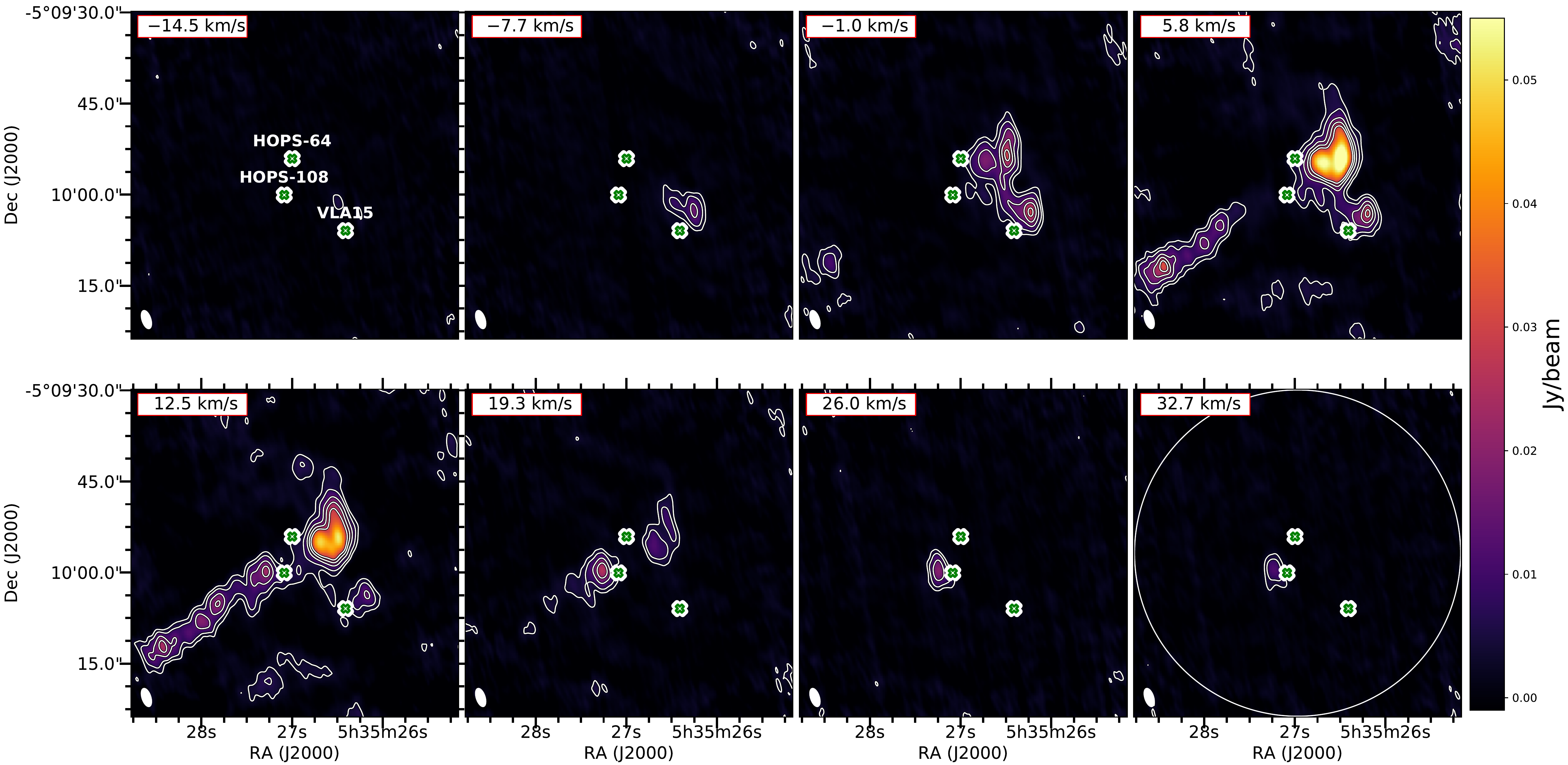}
    \caption{SiO (J\,=\,2-1) velocity channel map towards OMC-2 FIR4. White contours are -5 (dashed line), 10, 20, 30, and 40$\sigma$ levels, with a 1$\sigma=6.7\times10^{-4}$\,Jy/beam. The rms was derived for the whole map before the primary beam correction. The white-green crosses represent HOPS-108, HOPS-64, and VLA15 protostars in FIR4 (see Fig\,\ref{fig:SiO_mom0} caption for coordinates). The bottom left ellipses represent the synthesised beam, while the white open circle in the bottom right panel marks the primary beam FWHM for the NOEMA observations.}
    \label{fig:SiO_channels}
\end{figure*}

These new observations suggest the presence of monopolar outflows in FIR4, which could be more common than previously thought. \citet{2022A&A...667A...6C} showed a highly collimated ($\sim$\,1$^{\circ}$) monopolar SiO jet outflow originating from VLA15 in the south-west part of FIR4 (see Fig.\ref{fig:SiO_mom0}). Monopolar outflows are also found in low-mass objects \citep[e.g.][]{2014A&A...563L...3C} and among high-mass, star forming regions \citep[e.g.][]{2013ApJ...778...72F,2020A&A...636A..38N}. From a theoretical point of view, \citet{2018MNRAS.473.4868Z} recently revealed that asymmetric outflows can be more common than symmetric ones, due to the complexity of 3D structures during the process of protoplanetary disc formation and the fact that material infall (as well as the magnetic field geometry) is strongly asymmetric. \\
Figure\,\ref{fig:pv_plot} shows a position-velocity diagram obtained along a cut encompassing the SiO-collimated emission in the eastern part of FIR4 (see Fig.\,\ref{fig:SiO_mom0}). Despite the channel resolution of our data, the red emission (i.e. at higher offset in Fig.\,\ref{fig:pv_plot}, and hence closer to the HOPS-108 protostar) seems to be more spatially confined and reaches higher velocities than the blue emission. Similarly, in the channel maps displayed in Fig.\,\ref{fig:SiO_channels} the velocity components larger than 19 km/s are very compact, while the blue emission ($v_{lsr}$ < 13\,km/s), being projected in the plane-of-sky (more extended), has lower (line-of-sight) velocity components.\\

\begin{figure*}[htbp]
    \centering
    \includegraphics[scale=0.17]{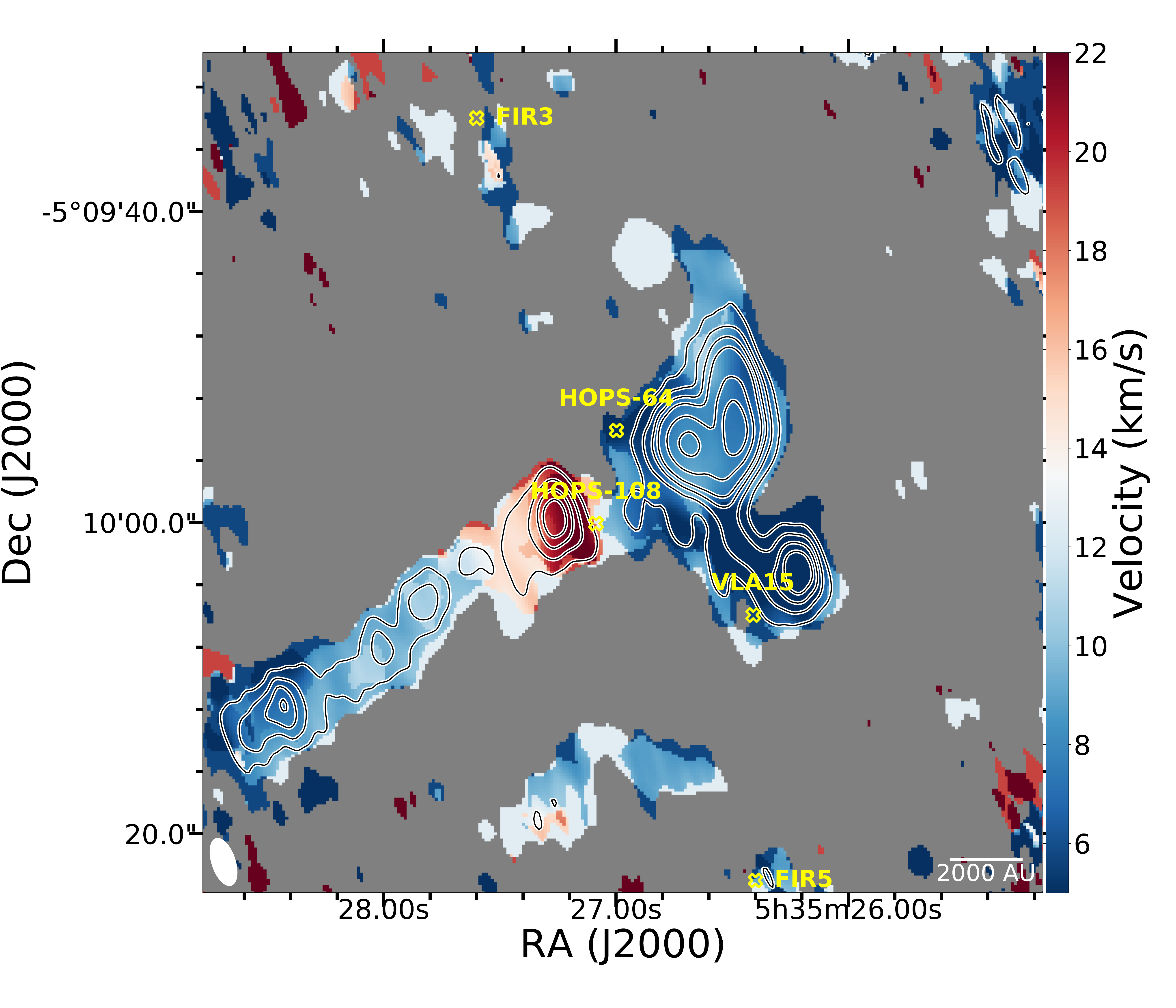}
    \includegraphics[scale=0.17]{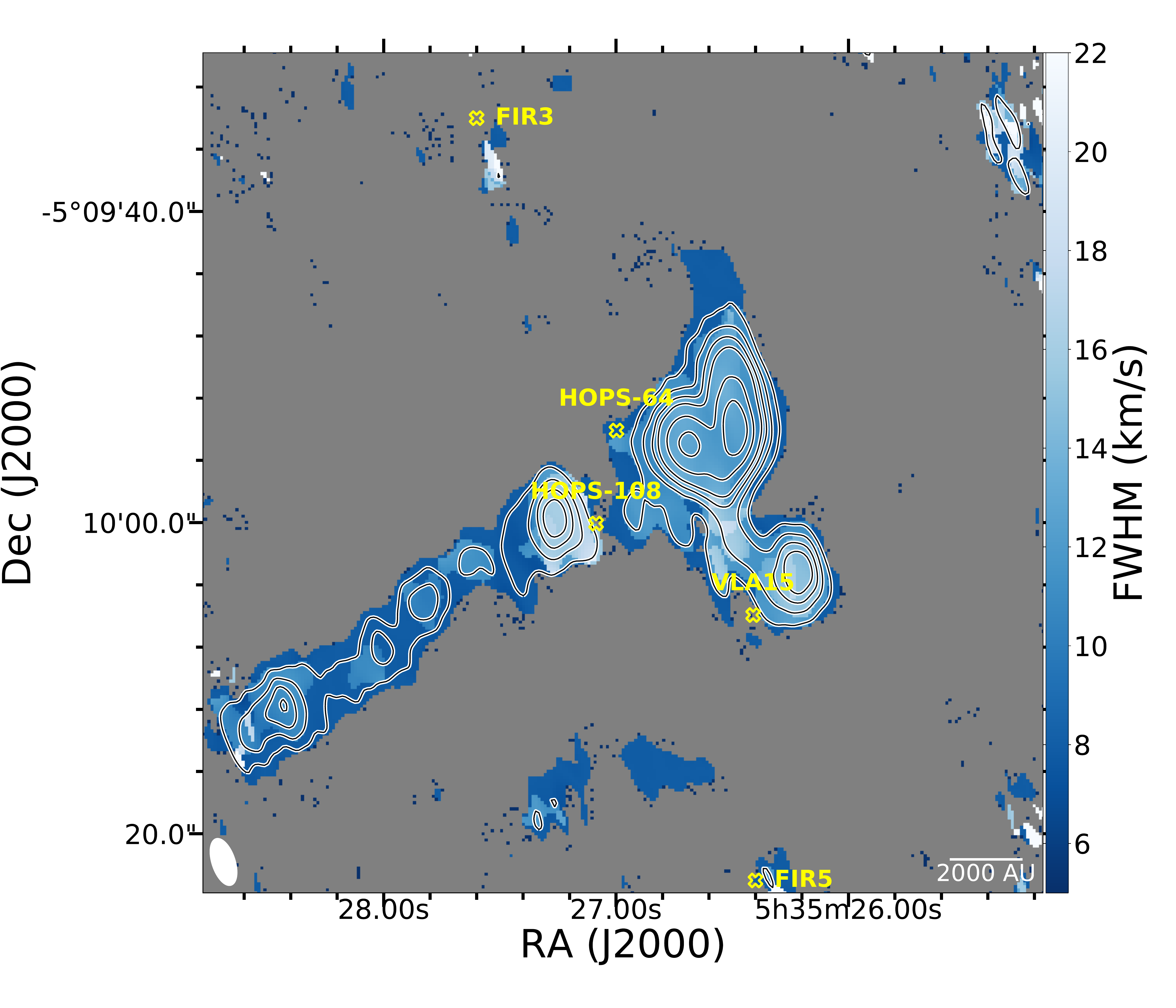}
    \caption{Moment 1 and 2 maps (upper and bottom panel, respectively) obtained with NOEMA observations. A 5$\sigma$ mask (with 1$\sigma=6.7\times10^{-4}$\,Jy/beam) was used for each map; white contours are the integrated intensity emissions, as presented in Fig.\,\ref{fig:SiO_mom0}. The bottom left white ellipses represent the synthesised beams. Light yellow labels indicate the main sources in the OMC-2 region.}
    \label{fig:moments}
\end{figure*}

The moment 1 map in Fig.\,\ref{fig:moments} shows the eastern region close to HOPS-108 embedded in a high-velocity ($\sim$\,16\,km\,s$^{-1}$) emission blob, its western counterpart with a lower and more homogeneous velocity distribution, and the protostar in between this two regimes. The increase in SiO (blue-shifted) velocities is followed by an increase in distance between the SiO emission and the protostar, which is evidence that the material is being accelerated by entrainment mechanisms \citep[e.g. jet-bow shock processes or decreasing cloud density gradient, see][]{2007prpl.conf..245A}. 

\subsection{Non-LTE modelling}\label{non-lte}

\begin{figure}[htbp]
    \includegraphics[width=\columnwidth]{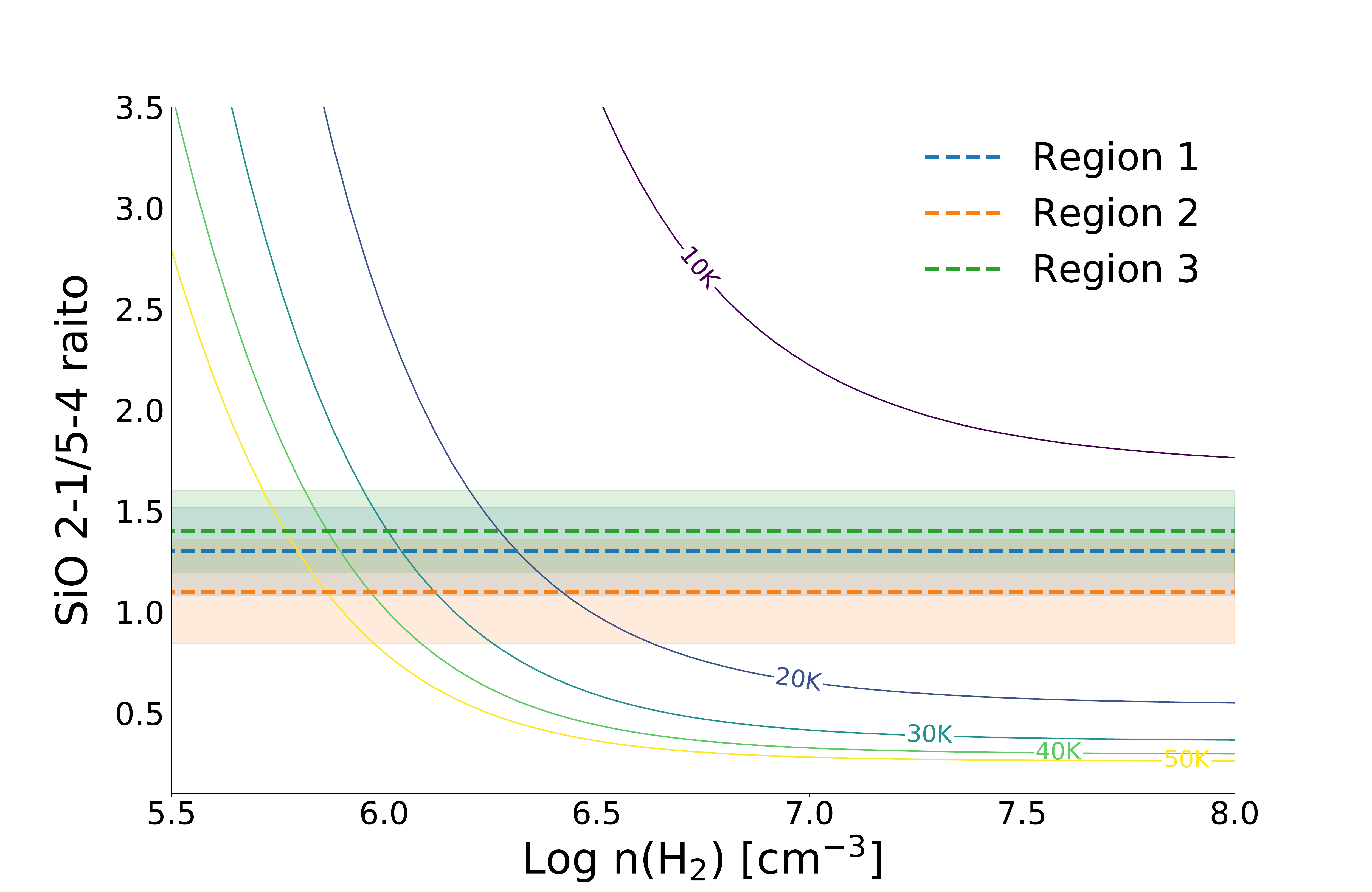}
    \caption{Non-LTE analysis of SiO emission lines. A series of RADEX models were run to estimate the kinetic temperature and density from the observed ratios of SiO $J$\,=\,2-1 and $J$\,=\,5-4 lines. Coloured dashed lines represent the ratio derived in the three regions highlighted in Fig.\,\ref{fig:SiO_mom0}; the coloured shaded regions correspond to the error bars of the derived ratios, after assuming a 20\% uncertainty on the peak brightness temperatures of each region. Solid curves represent the different temperatures obtained with the grid of models.}
    \label{fig:SiO_radex}
\end{figure}

A series of non-LTE RADEX\footnote{\url{ http://www.strw.leidenuniv.nl/~moldata/radex.html}}
\citep{2007A&A...468..627V} models were run to estimate the kinetic temperature and molecular hydrogen density from the observed ratios of the two SiO rotational transitions (Fig.\,\ref{fig:SiO_radex}). The input quantities for RADEX, namely the SiO column density (assuming optically thin emission) and the line width of the spectral lines, were estimated from Gaussian line fitting of the spectra performed with CASSIS\footnote{\url{http://cassis.irap.omp.eu/}} for each region. The average values were then used in the non-LTE analysis ($N({\rm SiO})=5\times10^{12}~$cm$^{-2}$ for the column density). Also, a kinetic temperature in the 10-80\,K range and a H$_2$ volume density in the range 10$^{4}$-10$^{8}$\,cm$^{-3}$ were used as boundary conditions for the model; these ranges were constrained by previous observations of the source and shock models. The ratio of the peak brightness temperatures of the two SiO lines ($J$\,=\,2-1 and 5-4), obtained from the three locations selected in the map, are shown by coloured dashed lines in Fig.\,\ref{fig:SiO_radex}, along with the different curves as a function of volume density and kinetic temperature. \\

\subsection{Modelling particle acceleration}

The origin of the high ionisation rate towards OMC-2 FIR4, revealed by \citet{Ceccarelli+2014} and confirmed by the subsequent studies of \citet{Fontani+2017} and \citet{Favre+2018}, has also been a subject of great interest from a theoretical perspective. \citet{Padovani+2015,Padovani+2016} and \citet{GachesOffner2018} pointed out that, in the shocks located along a protostellar jet or on the surface of a protostar, local acceleration of charged particles may take place according to the first-order Fermi acceleration mechanism. According to this process, thermal particles are accelerated to energies high enough to explain extreme high values of the ionisation rate as well as the synchrotron emission observed in jet knots that cannot be justified by only taking into account the average Galactic CR flux.

Based on recent continuum observations at millimetre and centimetre wavelengths \citep{Osorio+2017,Tobin+2019}, which established that FIR4 could fall in the path of the jet that originated in the HOPS-370 protostar (belonging to the FIR3 protostellar cluster), \citet{Padovani+2021} showed that an ionisation rate of $\approx4\times10^{-14}$~s$^{-1}$ can be reasonably expected if local particle acceleration occurs in the three knots of the HOPS-370 jet described in \citet{Osorio+2017}.

The discovery of a possible jet originating from the HOPS-108 protostar offers a distinct and appealing scenario for the local acceleration process, as the jet direction coincides with the side of the OMC-2 FIR4 protocluster where the high CR ionisation rate has been measured and spatially constrained. In addition, the central position of R1 with respect to the high-ionisation region (see Fig.\,\ref{fig:SiO_mom0}) and its proximity to the HOPS-108 protostar allows us to relax the model assumptions, namely the interaction between the FIR4 region and the HOPS-370 jet, which is still a matther of debate \citep{Favre+2018}.  We applied the model described in Sect.~2 of \citet{Padovani+2021} to R1 assuming a temperature of $10^4$~K \citep{Frank+2014} and a fully ionised medium at the shock front \citep{Araudo+2007}. These are typical assumptions in the case of intermediate- and high-mass protostellar jets. From the new observations, we obtained information on two key parameters for the acceleration process: the projected distance of R1 from the protostar (also known as shock radius, $R_{\rm sh}\sim1000$~AU) and the transverse size of the shock ($\ell_\perp\sim400$~AU). The other  parameters of the model, which are unknown, are the jet velocity in the shock  reference frame ($U$), the volume density ($n$), the fraction of ram pressure transferred to thermal particles ($\widetilde P$), and the magnetic field strength ($B$).

Following \citet{Padovani+2021}, we adopted a Bayesian method to infer the best-fit model, taking into account a set of values for each of the above parameters. In particular, we examined the following intervals: $50\le U/({\rm km~s^{-1}})\le1000$, $10^5\le n/{\rm cm^{-3}}\le10^9$, $10^{-6}\le\widetilde P\le10^{-2}$. 
We note that we considered the case of a parallel shock\footnote{A shock is parallel when the shock normal is parallel to the ambient magnetic field.}, which represents the simplest approach. In this case, the particle acceleration timescale turns out to be independent of the magnetic field strength\footnote{See Eqs.~(1$-$3) in \citet{Padovani+2021}.}, so 
we would need synchrotron multi-frequency observations as in the case presented by \citet{Padovani+2021} to constrain $B$.

For each combination of the parameters, we calculated the flux of accelerated protons\footnote{The proton flux is defined as the number of protons per unit of energy, time, area, and solid angle.} at the shock surface of R1, $j_p^{\rm sh}$. Then, we computed the propagation of the proton flux in each shell of radius $r$ from $\ell_\perp/2=200$~AU to $R_{\rm ion}\simeq5000$~AU, which is the average radius of the region where \citet{Fontani+2017} found $\zeta=4\times10^{-14}$~s$^{-1}$ (see Fig.\,\ref{fig:SiO_mom0}). Accounting for the attenuation of the flux according to the continuous slowing-down approximation, the proton flux in each shell is given by
\be\label{shellflux}
j_{p}(E,r,\delta)=j_{p}^{\rm sh}(E_{0})\frac{L(E_{0})}{L(E)}%
        \left(\frac{{\ell_\perp/2}}{{\ell_\perp/2}+r}\right)^{-\delta}\,,
\ee
where 
$L$ is the proton energy-loss function \citep{Padovani+2009}, and $E$ is the energy of a proton with initial energy $E_0$ after passing through a column density $N=nr$. The $\delta$ parameter models the propagation of the protons and their relative energy loss depending on the environmental conditions; the two limiting cases are of pure free-streaming (i.e. geometrical dilution, $\delta\,=\,2$)  and when the propagation is attenuated because of diffusion \citep[$\delta\,=\,1$;][]{Aharonian2004}.
Finally, we computed the mean proton flux averaging over the volume of the spherical shell,
\be\label{averagejp}
\langle j_{p}(E,\delta)\rangle=\left(\frac{4\pi}{3} R_{\rm ion}^{3}\right)^{-1}%
        \int_{0}^{R_{\rm ion}}4\pi r^{2}j_{p}(E,r,\delta)\ud r\,.
\ee
The corresponding ionisation rate is
\be\label{zetaion}
\zeta_{\delta}=2\pi\int\langle j_{p}(E,\delta)\rangle\sigma_{\rm ion}(E)\ud E\,,
\ee
where $\sigma_{\rm ion}$ is the ionisation cross-section for protons colliding with molecular hydrogen \citep[see e.g.][]{Rudd+1992}. We then proceeded as follows: for each set of ($U,n,\widetilde P$), we computed the expected ionisation rate in the case of diffusion and geometrical dilution ($\zeta_1$ and $\zeta_2$, respectively). In the case that the ionisation rate estimated from the observations, $\zeta$, falls in the interval of $[\zeta_2,\zeta_1]$, we assumed that the data were characterised by Gaussian uncertainties, so the likelihood of a given model is proportional to $\exp(-\chi^2_\delta/2)$, with $\chi^2_\delta=(\zeta-\zeta_\delta)^2/\sigma_\zeta^2$, where we assumed $\sigma_\zeta/\zeta=25\%$.

\begin{figure}[htbp]
    \centering
    \includegraphics[width=\columnwidth]{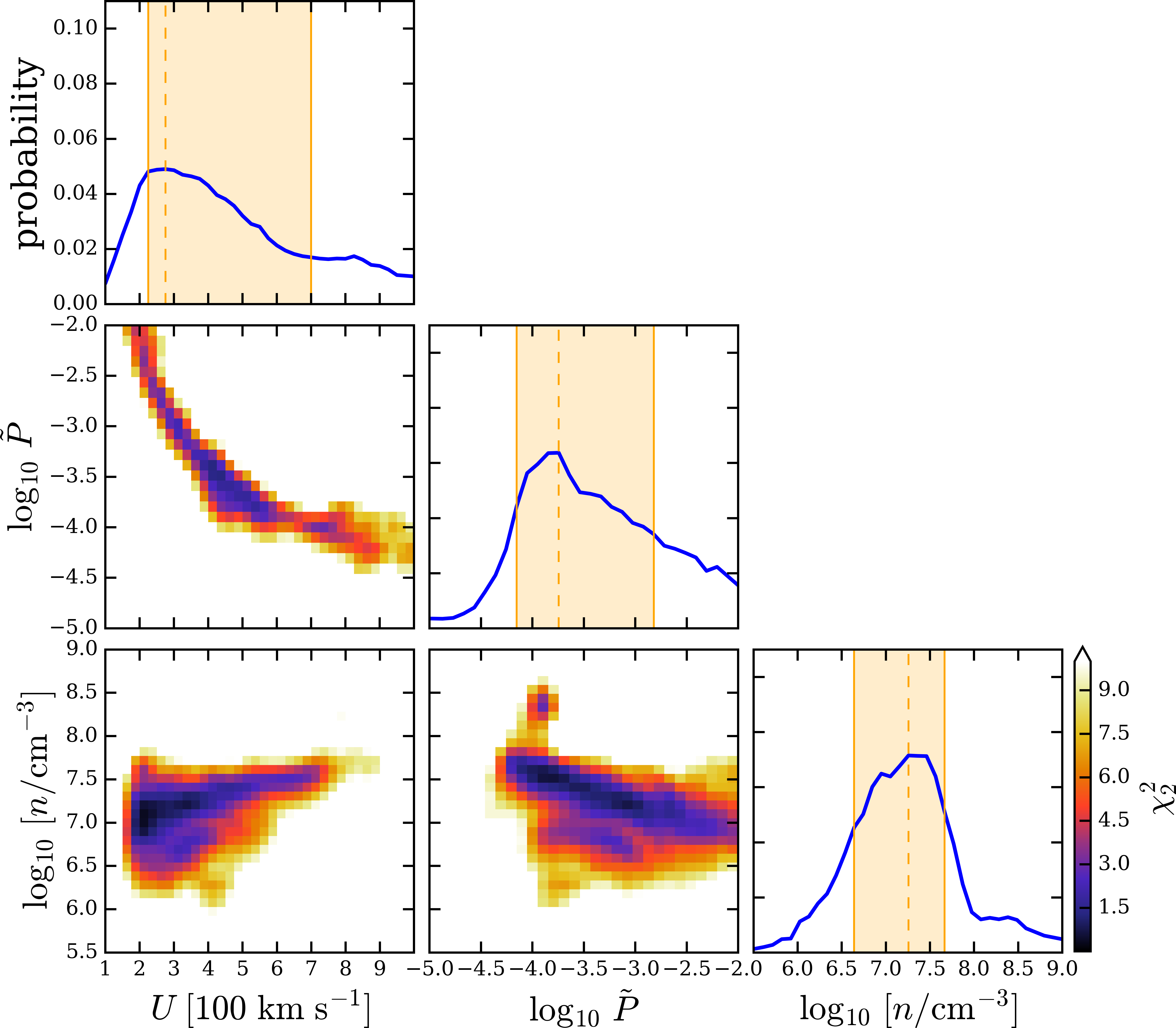}
    \caption{Corner plot of $\chi_\delta^2$ surface as a function of the model parameters for the case of pure geometrical dilution
    ($\delta=2$). {\em Top panels} in each column: Probability density distributions for the marginalised parameters; confidence intervals (first and third quartiles) are shown as orange shaded rectangular regions, and the maximum-likelihood estimate is shown by a vertical dashed line.} 
    \label{fig:cornerplot}
\end{figure}

Since the best-fit parameter values computed in the diffusion and the geometrical dilution cases vary by less than 15\%, we only discuss the results for the pure geometrical dilution case ($\chi^2_2$). Fig.\,\ref{fig:cornerplot} shows the corner plot of the best fit: the quantities $U$, $n$, and $\widetilde P$ show clear correlations and their probability distributions all show a rather pronounced peak (errors are estimated using the first and third quartiles). The best-fit ranges are $U = 275_{-50}^{+425}~{\rm km\ s^{-1}}$, $n/10^6 = 18.05_{-13.67}^{+28.37}~{\rm cm^{-3}}$, and $\widetilde P/10^{-4} = 1.80_{-1.10}^{+13.31}$. Clearly, the errors are rather large, and this is due to the fact that we only have two observational constraints (the projected distance of R1 from HOPS-108 and the transversal size of the shock). However, the central values are consistent with those expected in this type of region  \citep{Padovani+2021,Araudo+2021}.

\section{Discussion}\label{sec:discussion}

In the last ten years, several works focused on OMC-2 FIR4 to try to shed some light on the complexity of this protocluster region. Through analysis of the interaction between FIR3 and FIR4 via the [O I] jet originating from the former, \citet{Gonzalez-Garcia+_2016} also pointed out that the peak of the [O I] emission, along with all the other high-excitation molecular cooling lines observed in the far-IR (H$_2$O, CO, and OH) with \textit{Herschel}/PACS, has its emission peak at the location of FIR4 \citep{Furlan+_2014}. The lack of a jet outflow from FIR4 led the authors to the conclusion that the bright line [O I] emission seen towards FIR4 is the terminal shock (Mach disc) of the FIR3 jet. On the other hand, they also stated that FIR4 may simply lie along the line of sight and may not even be physically associated with the shocked emitting region. Similar conclusions concerning the interaction of FIR3 and FIR4 were proposed by \citet{Osorio+2017}, although the large proper motion velocity of HOPS-108 was inconsistent with the triggered scenario. The authors suggested that, alternatively, an apparent proper motion could result because of a change in the position of the centroid of the source due to a one-sided eject of ionised plasma, rather than by the actual motion of the protostar itself. 

Previous SOLIS observations of this source revealed spatial variation among distinct carbon-chains towards FIR4. In \citet{Fontani+2017}, the whole FIR4 source was divided into two sub-regions, following the ratio of HC$_3$N/HC$_5$N emission \citep[see Fig.\,1 of\,][]{Fontani+2017}. While in the western part this ratio is high (10-30), the eastern region of FIR4, identified by the green contour in Fig. 1, is rich in HC$_5$N, making the HC$_3$N/HC$_5$N ratio in the 4-12 range. The authors interpreted this variation as the result of an enhancement of $\sim$\,1000 in the CR ionisation rate (with respect to the value deduced in molecular clouds) towards the eastern side of FIR4. In particular, the eastern half was proposed to be strongly irradiated, while the western region partially shielded. The SiO maps presented in this work are in agreement with this hypothesis. The emission morphology of SiO across the FIR4 region clearly shows a collimated stream towards the south-east, and a more compact blob in the western region. The jet candidate could in fact produce the acceleration of energetic particles as modelled by \citet{Padovani+2016}. This finding is also consistent with the detection of bright free-free emission partially overlapping with the FIR4-HC$_5$N region observed in \citet{Fontani+2017} and extending outside the eastern border of FIR4 \citep{1999AJ....118..983R}. In Fig.\,\ref{fig:SiO_mom0}, the SiO jet emission coincides with the FIR4-HC$_5$N emission reported by \citet{Fontani+2017}. From this comparison, it is clear how HC$_5$N and the SiO-jet trace the same, eastern part of FIR4.

\citet{Tobin+2019} reported highly compact methanol emission originating from $\sim$\,100\,AU scale coincident with this source. Although thermal evaporation of ices due to the thermal dust heating produced by the nearby HOPS-108 was proposed by the authors as the simplest explanation for the observed methanol emission, they also stated that shock heating might explain the chemical richness in molecular lines observed towards HOPS-108. This also suggests that a jet driven by HOPS-108 may be present. A possible scenario would be that a protostellar jet and the highly energetic ionised plasma accelerated by the protostar impacts the material nearby, producing shocked gas traced by the SiO.
Using this assumption, the model implemented in this work shows that the conditions of the shock regions traced by SiO, in particular close to the protostar HOPS-108 (i.e. R1), are favourable for the acceleration of thermal particles, boosting their energies high enough to explain the ionisation rate inferred by previous studies. Previous theoretical works had already shown that the high ionisation rate observed in the FIR4 region can be explained by the presence of cosmic rays locally accelerated on the protostellar surface \citep{Padovani+2016}, in a protostellar cluster \citep{GachesOffner2018,2019ApJ...878..105G}, or in jet shocks \citep{Padovani+2021}. Thanks to the new high angular-resolution observations presented in this article, we have been able to identify the local CR source much more precisely. With all the limitations due to the current dataset, the model indeed derives physical-chemical parameters in accordance with those obtained by previous works. In particular, the H$_2$ volume density is in good agreement with the values obtained using different techniques \citep[e.g.][]{Ceccarelli+2014,2013A&A...556A..62L}. Similar considerations apply to the results of the non-LTE analysis of the SiO emission spectra extracted from the regions along the jet itself. The values of the kinetic temperature and density obtained by modelling the ratio of the observed transitions are consistent with those obtained by previous studies (20\,K < T$_{kin}$< 50\,K for densities $\sim 5\times10^6$cm$^{-3}$).\\

\section{Conclusions}

New NOEMA observations allowed us to shed some light on the protocluster FIR4 in the OMC-2 region, and to propose a new scenario to explain its puzzling high ionisation rate. The main outcomes of the present study are as follows:

\begin{enumerate}
    \item The detection of a jet candidate originating from within the FIR4 cloud for the first time, which was seen in the SiO (2-1) line.
    \item The jet candidate is extending towards the east side of OMC-2 FIR4, in the same region where a high ionisation rate was previously measured. 
    \item Our observations suggest that the protostellar HOPS-108 might be the driving source candidate of the SiO jet.
    \item Modelling the acceleration of particles along the collimated emission, we show that the high ionisation can indeed be produced by the newly discovered jet driven by HOPS-108.
\end{enumerate}

This acceleration allows the particle to gain enough energy to explain the ionisation rate inferred in the region by previous studies. Future analyses on this collimated emission, performed with higher angular and spectral resolution, might help to further elucidate the complex kinematics of this system.

\begin{acknowledgements}

This project has received funding within the European Union’s Horizon 2020 research and innovation programme from the European Research Council (ERC) for the project “The Dawn of Organic Chemistry” (DOC), grant agreement No 741002, and from the Marie Sklodowska-Curie for the project ”Astro-Chemical Origins” (ACO), grant agreement No 811312. This paper makes use of the following ALMA data: ADS/JAO.ALMA\#2017.1.01353.S. ALMA is a partnership of ESO (representing its member states), NSF (USA) and NINS (Japan), together with NRC (Canada), MOST and ASIAA (Tai- wan), and KASI (Republic of Korea), in cooperation with the Republic of Chile. The Joint ALMA Observatory is operated by ESO, AUI/NRAO and NAOJ. V.L., F.O.A., and P.C. acknowledge financial support from the Max Planck Society. We thank Alexei Ivlev and Jaime E. Pineda for useful discussions.

\end{acknowledgements}

\bibliographystyle{aa}
\bibliography{OMC2_SOLIS_arxiv}

\end{document}